\documentclass[10pt, preprint]{sigplanconf}

\usepackage{epsfig}
\usepackage{boxedminipage}
\usepackage{latexsym}
\usepackage{balance}

\usepackage{amsmath}
\usepackage{amsthm}
\usepackage{amssymb}
\usepackage{hyperref}
\usepackage{graphicx}
\usepackage{subfigure}
\usepackage[english]{babel}
\usepackage[utf8x]{inputenc}
\usepackage{eso-pic}
\usepackage{alltt}
\usepackage{listings}
\usepackage{algorithm}
\usepackage{algorithmic}
\usepackage[font=small]{caption}
\usepackage{framed}
\usepackage{float}
\usepackage{subfloat}
\usepackage{wrapfig}
\usepackage{multirow}

\def\rem#1{}
\def\fullpaper#1{}

\newcommand{\kipf}{\mbox{$k$-IPF}}
\newcommand{\ksf}{\mbox{$k$-SF}}
\newcommand{\kblpp}{\mbox{$k$-BLPP}}
\newcommand{\blpp}{\mbox{BLPP}}

\newtheorem{definition}{Definition}

\begin{document}

\conferenceinfo{WXYZ '13}{date, City.} 
\copyrightyear{2013} 
\copyrightdata{[to be supplied]} 

\newcommand{\emilio}{{$\bullet \bullet \bullet $}}
\newcommand{\camil}{{$\bullet \bullet \bullet $}}
\newcommand{\irene}{{$\bullet \bullet \bullet $}}
\newcommand{\romolo}{{$\bullet \bullet \bullet $}}

\title{Ball-Larus Path Profiling Across Multiple Loop iterations}

\authorinfo{Daniele Cono D'Elia}
           {{\small Dept. of Computer, Control and Management Engineering\\Sapienza University of Rome}}
           {delia@dis.uniroma1.it}
\authorinfo{Camil Demetrescu}
           {{\small Dept. of Computer, Control and Management Engineering\\Sapienza University of Rome}}
           {demetres@dis.uniroma1.it}
\authorinfo{Irene Finocchi}
           {{\small Dept. of Computer Science\\Sapienza University of Rome}}
           {finocchi@di.uniroma1.it}


\maketitle

\begin{abstract}
Identifying the hottest paths in the control flow graph of a routine can direct optimizations to portions of the code where most resources are consumed. This powerful methodology, called {\em path profiling}, was introduced by Ball and Larus in the mid 90's~\cite{BL96} and has received considerable attention in the last 15 years for its practical relevance. A shortcoming of Ball-Larus path profiling was the inability to profile cyclic paths, making it difficult to mine interesting execution patterns that span multiple loop iterations. Previous results, based on rather complex algorithms, have attempted to circumvent this limitation at the price of significant performance losses already for a small number of iterations.  In this paper, we present a new approach to multiple-iterations path profiling, based on data structures  built on top of the original Ball-Larus numbering technique. Our approach allows it to profile {\em all executed paths} obtained as a concatenation of up to $k$ Ball-Larus acyclic paths, where $k$ is a user-defined parameter. 
An extensive experimental investigation on a large variety of Java benchmarks on the Jikes RVM shows that, surprisingly, our approach can be even faster than Ball-Larus due to fewer operations on smaller hash tables, producing compact representations of cyclic paths even for large values of $k$.



\end{abstract}



\section{Introduction}
\label{se:intro}

Path profiling is a powerful methology for identifying performance bottlenecks in a program. The approach consists of associating performance metrics, usually frequency counters, to paths in the control flow graph. Identifying hot paths can direct optimizations to portions of the code that could yield significant speedups. For instance, trace scheduling can improve performance by increasing instruction-level parallelism along frequently executed paths~\cite{Fisher:1981:TST:1311075.1311325}. The seminal paper by Ball and Larus~\cite{BL96} introduced a simple and elegant path profiling technique. The main idea was to implicitly number all possible acyclic paths in the control flow graph so that each path is associated with a unique compact path identifier (ID). The authors showed that path IDs can be efficiently generated at runtime and can be used to update a table of frequency counters. Although in general the number of acyclic paths may grow esponentially with the graph size, in typical control flow graphs this number is usually small enough to fit in current machine wordsizes, making this approach very effective in practice. 

While the original Ball-Larus approach was restricted to acyclic paths obtained by cutting paths at loop back edges, profiling paths that span consecutive loop iterations is a desirable yet difficult task that can yield better optimization opportunities. Consider for instance the problem of eliminating redundant executions of instructions, such as loads and stores~\cite{Bodik:1999:LAD:301618.301643}, conditional jumps~\cite{Bodik:1997:ICB:258915.258929}, expressions~\cite{Bodik:2004:CRR:989393.989453}, and array bounds checks~\cite{Bodik:2000:AEA:349299.349342}. A typical situation is that the same instruction is redundantly executed at each loop iteration, which is particularly common for arithmetic expressions and load operations~\cite{Bodik:2004:CRR:989393.989453, Bodik:1999:LAD:301618.301643}. To identify such redundancies, paths that extend across loop back edges need to be profiled. Another application is trace scheduling~\cite{Fisher:1981:TST:1311075.1311325}: if a frequently executed cyclic path is found, compilers may unroll the loop and perform trace scheduling on the unrolled portion of code. Tallam {\em et al.}~\cite{Tallam:2004:EPP:977395.977659} provide a comprehensive discussion of the benefits of multi-iterations path profiling. 


Different authors have proposed techniques to profile cyclic paths by modifying the original Ball-Larus path numbering scheme in order to identify paths that extend across multiple loop iterations~\cite{Tallam:2004:EPP:977395.977659,SS09,Li20121558}. Unfortunately, all known solutions require rather complex algorithms that incur severe performance overheads even for short cyclic paths, leaving it as an interesting open question to find simpler and more efficient alternative methods.


\paragraph{Our results.} In this paper, we present a novel approach to multiple-iterations path profiling, which provides substantially better performance than previous techniques even for long paths. Our method stems from the observation that any cyclic execution path in the control flow graph of a routine can be described as a concatenation of Ball-Larus acyclic paths (BL paths). In particular, we show how to accurately profile {\em all executed paths} obtained as a concatenation of up to $k$ BL paths, where $k$ is a user-defined parameter. To do so, we reduce multiple-iterations path profiling to the problem of counting $n$-grams, i.e., contiguous sequences of $n$ items from a given sequence. To compactly represent collected profiles, we organize them in a prefix tree (or trie)~\cite{Fredkin60} of depth up to $k$ where each node is labeled with a BL path, and paths in the tree represent concatenations of BL paths that were actually executed by the program, along with their frequencies. 


We implemented our ideas by developing a Java performance profiler in the Jikes Research Virtual Machine~\cite{Jikes}. To make fair performance comparisons with state-of-the-art previous profilers, we built our code on top of the BLPP profiler developed by Bond~\cite{BM05,JikesResearchArchive}, which provides an efficient implementation of the Ball-Larus acyclic path profiling technique.\rem{other ref?} A broad experimental study on a large suite of prominent Java benchmarks on the Jikes Research Virtual Machine shows that our profiler can trace long paths efficiently, making it possible to collect profiles that would have been too costly to gather using previous multi-iterations techniques.

\begin{figure}
\centerline{\includegraphics[width=0.7\columnwidth]{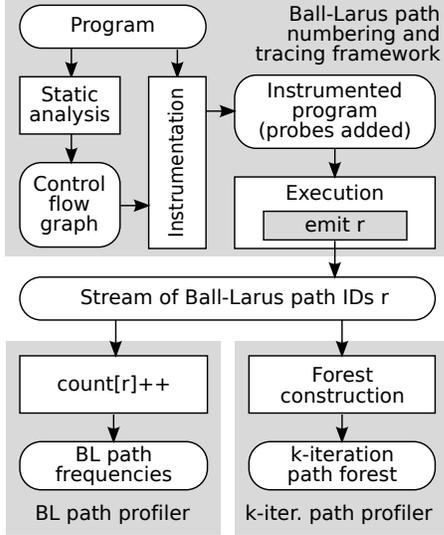}}
\medskip
\caption{Overview of our approach: classical Ball-Larus profiling versus k-iteration path profiling, cast in a common framework.}
\label{fig:approach}
\end{figure}

\paragraph{Techniques.} Differently from previous approaches~\cite{Tallam:2004:EPP:977395.977659,SS09,Li20121558}, which rely on modifying the Ball-Larus path numbering to cope with cycles, our method does not require any modification of the original numbering technique described in~\cite{BL96}. 
The main idea behind our approach is to fully decouple the task of tracing Ball-Larus acyclic paths at run time from the task of concatenating and storing them in a data structure to keep track of multiple iterations. The decoupling is performed by letting the Ball-Larus profiling algorithm issue a stream of BL path IDs (see Figure~\ref{fig:approach}), where each ID is generated when a back edge in the control flow graph is traversed or the current procedure is abandoned. As a consequence of this modular approach, our method can be implemented on top of existing Ball-Larus path profilers, making it simpler to code and maintain. 



Our profiler introduces a technical shift based on a smooth blend of the path numbering methods used in intraprocedural path profiling with data structure-based techniques typically adopted in interprocedural profiling, such as calling context profiling. The key to the efficiency of our approach is to replace costly hash table accesses, which are required by the Ball-Larus algorithm to maintain path counters for non-small programs, with substantially faster operations on trees. With this idea, we can profile paths that extend across many loop iterations in comparable time, if not faster, than profiling acyclic paths on a large variety of industry-strength benchmarks.

\paragraph{Organization of the paper.} In Section~\ref{se:approach} we describe our approach and 
in Section~\ref{se:implementation} we discuss how to implement it. The results of our experimental investigation are detailed in Section~\ref{se:experiments} and related work is surveyed in Section~\ref{se:related}. Concluding remarks are given in Section~\ref{se:conclusion}.



\begin{figure}[t]
\begin{small}
\noindent {\bf procedure} {\tt bl\_path\_numbering}$()$:
\begin{algorithmic}[1]
\FOR {each basic block $v$ in reverse topological order}
\IF {$v$ is the exit block}
\STATE numPaths($v$) $\gets 1$
\ELSE
\STATE numPaths($v$) $\gets 0$
\FOR {each outgoing edge $e = (v, w)$}
\STATE val($e$) = numPaths($v$)
\STATE numPaths($v$) += numPaths($w$)
\ENDFOR
\ENDIF 
\ENDFOR
\end{algorithmic}

\end{small}
\smallskip
\caption{Ball-Larus path numbering algorithm.}
\label{fig:bl-numbering}
\end{figure}

\begin{figure*}[t]
\centerline{\includegraphics[width=0.95\textwidth]{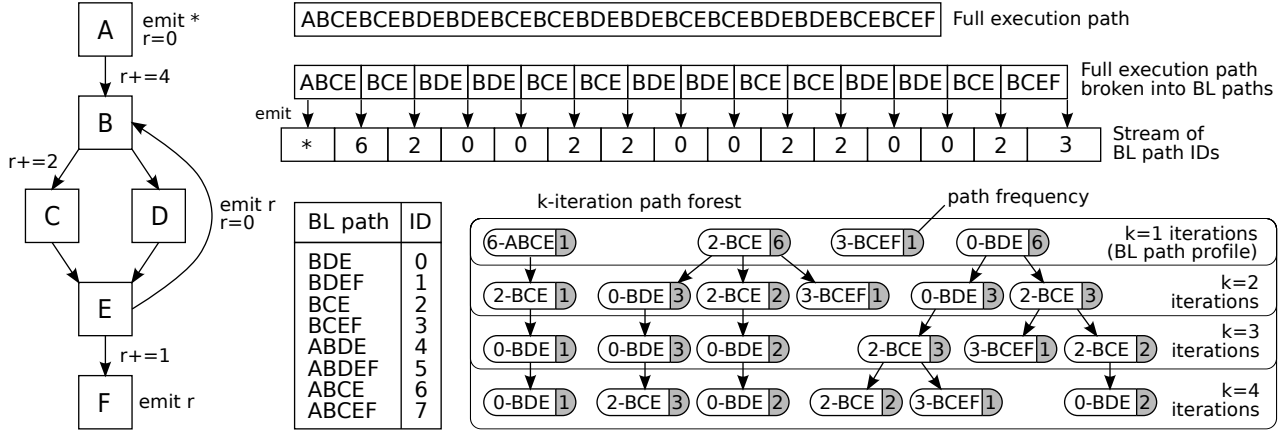}}
\medskip
\caption{Control flow graph with Ball-Larus instrumentation modified to emit acyclic path IDs to an output stream and running example of our approach that shows a 4-iteration path forest (4-IPF) for a possible small execution trace.}
\label{fig:example}
\end{figure*}

\section{Approach}
\label{se:approach}

In this section we provide an overview of our approach to multiple-iterations path profiling. From a high level point of view, illustrated in Figure~\ref{fig:approach}, the entire process is divided into two main phases: 

\begin{enumerate}
\item instrumentation and execution of the program to be profiled (top of Figure~\ref{fig:approach});
\item profiling of paths (bottom of Figure~\ref{fig:approach}).
\end{enumerate}

\noindent The first phase is almost identical to the original approach described in~\cite{BL96}. The target program is statically analyzed and a control flow graph (CFG) is constructed for each routine of interest. The CFG is used to instrument the original program by inserting probes, which allow it to trace paths at run time. When the program is executed, taken acyclic paths are identified using the inserted probes. The main difference with the Ball-Larus approach is that, instead of directly updating a frequency counters table here, we emit a stream of path IDs, which is passed along to the next stage of the process. This allows us to decouple the task of tracing taken paths from the task of profiling them. 

The profiling phase can be either the original hash table-based method of~\cite{BL96} used to maintain BL path frequencies (bottom-left of Figure~\ref{fig:approach}), or other approaches such as the one we propose, i.e., profiling concatenations of BL paths in a forest-based data structure (bottom-right of Figure~\ref{fig:approach}). Different profiling methods can  be therefore cast into a common framework, increasing flexibility and helping us make more accurate comparisons.

We start with a brief overview of the Ball-Larus path tracing technique, which we use as a the first stage of our profiler.

\subsection{Ball-Larus Path Tracing Algorithm}

The Ball-Larus path profiling (BLPP) technique~\cite{BL96} identifies each acyclic path that is executed in a routine. Paths start on the method entry and terminate on the method exit. Since loops make the CFG cyclic, loop back edges are substituted by a pair of dummy edges: the first one from the method entry to the target of the loop back edge, and the second one from the source of the loop back edge to the method exit. After this (reversible) transformation, the CFG of a method becomes a DAG (directed acyclic graph) and acyclic paths can be enumerated.

The Ball-Larus path numbering algorithm, shown in Figure~\ref{fig:bl-numbering}, assigns a value $val(e)$ to each edge $e$ of the CFG such that, given N acyclic paths, the sum of the edge values along any entry-to-exit path is a unique numeric ID in [0, N-1]. A CFG example and the corresponding path IDs are shown in Figure~\ref{fig:example}: notice that there are eight distinct acyclic paths, numbered from 0 to 7, starting either on the method's entry $A$, or at loop header $B$ (target of back edge $(E,B)$).

BLPP places instrumentation on edges to compute a unique path number for each possible path. In particular, it uses a variable {\tt r}, called {\em probe} or {\em path register}, to compute the path number. Variable {\tt r} is first initialized to zero upon method entry and then is updated as edges are traversed. When an edge that reaches the method exit is executed, or a back edge is traversed, variable {\tt r} represents the unique ID of the taken path. As observed, instead of using the path ID {\tt r} to increase the path frequency counter ({\tt count[r]++}), we defer the profiling stage by emitting the path ID to an output stream ({\tt emit r}). To support profiling over multiple invocations of the same routine, we annotate the stream with the special marker $*$ to denote a routine entry event. Instrumentation code for our CFG example is shown on the left of Figure~\ref{fig:example}.



\subsection{$k$-Iterations Path Profling}
\label{sse:kIterationsPP}

The second stage of our profiler takes as input the stream of BL path IDs generated by the first stage and uses it to build a data structure that keeps track of the frequencies of {\em each and every distinct taken path} consisting of the concatenation of up to $k$ BL paths, where $k$ is a user-defined parameter. This problem is equivalent to counting all $n$-grams, i.e., contiguous sequences of $n$ items from a given sequence of items, for each $n\le k$. Our solution is based on the notion of {\em prefix forest}, which compactly encodes a list of sequences by representing only once repetitions and common prefixes. A prefix forest can be defined as follows:

\begin{definition}[prefix forest]
Let $L=\langle x_1, x_2, \ldots, x_q\rangle$ be any list of finite-length sequences over an alphabet $H$. A {\em prefix forest} ${\cal F}(L)$ of $L$ is a minimal labeled forest such that, for each $x_i=\langle a_1, a_2, \ldots, a_n\rangle\in L$ there is a path $\pi_i=\langle \alpha_1, \alpha_2, \ldots, \alpha_n\rangle\in {\cal F}(L)$ where each node $\alpha_j$, $j\in[1,n]:$
\begin{enumerate}
\item is labeled with $a_j$, i.e., $\ell(\alpha_j)=a_j\in H$;
\item has an associated counter $c(\alpha_j)$ that counts the number of times $\langle a_1, a_2, \ldots, a_j\rangle\subseteq x_i$ occurs in $L$.
\end{enumerate}
\end{definition}



\paragraph{k-Iterations Path Forest.} The output of our profiler is a prefix forest, which we call {\em k-Iterations Path Forest} (\kipf), that compactly represents all observed contiguous sequences of up to $k$ BL path IDs:

\begin{definition}[$k$-Iterations Path Forest]
\label{de:kipf}
Given an input stream $\Sigma$ representing a sequence of BL path IDs and $*$ markers, the {\em k-Iterations Path Forest} (\kipf) of $\Sigma$ is defined as $\kipf={\cal F}($list of all $n$-grams of $\Sigma$ that do not contain $*$, with $n\le k)$.
\end{definition}

\noindent By Definition~\ref{de:kipf}, the \kipf\ is the prefix forest of all consecutive subsequences of up to $k$ BL path IDs in $\Sigma$.


\paragraph{Example 1.} Figure~\ref{fig:example} provides an example showing the 4-IPF constructed for a small sample execution trace consisting of a sequence of 44 basic blocks encountered during one invocation of the routine described by the control flow graph on the left. Notice that the full (cyclic) execution path starts from the entry basic block $A$ and terminates on the exit basic block $F$. The first stage of our profiler issues a stream $\Sigma$ of BL path IDs that are obtained by emitting the value of the probe register $r$ each time a back edge is traversed, or the exit basic block is executed. Observe that the sequence of emitted path IDs induces a partition of the execution path into Ball-Larus acyclic paths. Hence, the sequence of executed basic blocks can be fully reconstructed from the sequence $\Sigma$ of path IDs.

The 4-IPF built in the second stage contains exactly one tree for each of the 4 distinct BL path IDs (0, 2, 3, 6) that occur in the stream. Notice that path frequencies in the first level of the 4-IPF are exactly those that traditional Ball-Larus profiling would collect. The second level contains the frequencies of taken paths obtained by concatenating 2 BL paths, etc. Notice that the path labeled with $\langle 2, 0, 0, 2\rangle$ in the 4-IPF, which corresponds to the path $\langle B, C, E, B, D, E, B, D, E, B, C, E \rangle$ in the control flow graph, is a 4-gram that occurs 3 times in $\Sigma$ and is one of the most frequent paths among those that span from 2 up to 4 loop iterations.

\paragraph{Properties.} 
A \kipf\ has some relevant properties:
\begin{enumerate}
\item $\forall$ node $\alpha\in$ $k$-IPF, $k>0$: $$c(\alpha)\ge\sum_{\beta_i\,:\,(\alpha,\beta_i)\in k\mbox{\scriptsize-IPF}} c(\beta_i);$$
\item $\forall k>0$, $k$-IPF $\subseteq$ ($k+1$)-IPF.
\end{enumerate}

\noindent By Property~1, since path counters are non-negative, they are monotonically non-increasing as we walk down the tree. The inequality $\ge$ in Property~1 may be strict ($>$) if the execution trace of a routine invocation does not end at the exit basic block; this may be the case when a subroutine call is performed at an internal node of the CFG. Notice that a 1-IPF includes only acyclic paths and yields exactly the same counters as a Ball-Larus profiler~\cite{BL96}.

\begin{figure}
\centerline{\includegraphics[width=0.99\columnwidth]{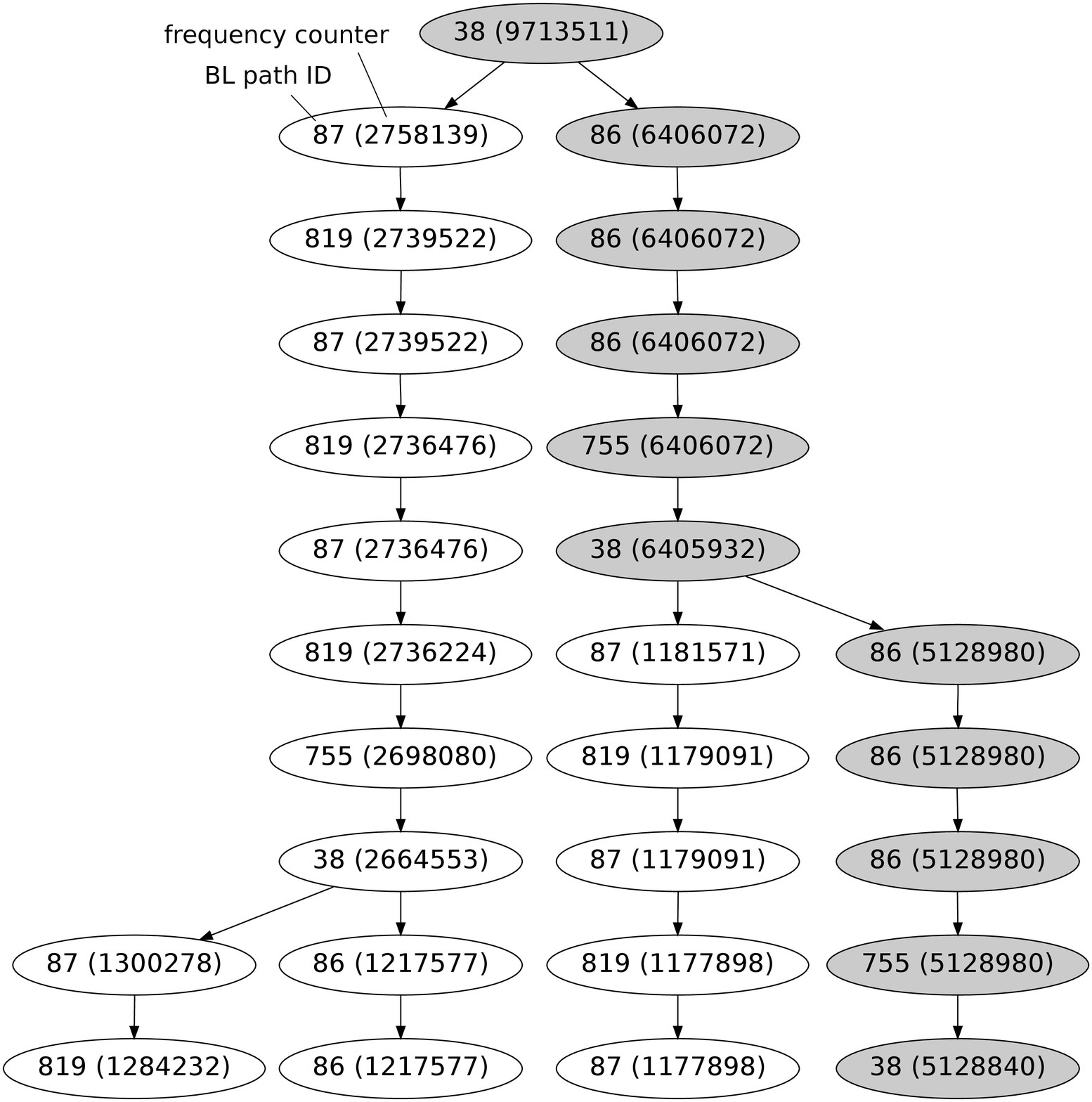}}
\caption{Subtree of the $11$-IPF of method {\tt org.eclipse.jdt.} {\tt internal.compiler.parser.Scanner.checkTaskTag} taken from release 2006-MR2 of the {\tt DaCapo} benchmark suite. 
}
\label{fig:eclipse-example}
\end{figure}

\paragraph{Example 2.} In Figure~\ref{fig:eclipse-example} we show a subtree of the $11$-IPF generated for method {\tt checkTaskTag} of class {\tt Scanner} in the {\tt org.eclipse.jdt.internal.} {\tt compiler.parser} package of the {\tt eclipse} benchmark included in the {\tt DaCapo} release 2006-MR2. In the subtree, we pruned all nodes with counters less than 10\% of the counter of the root. Notice that, after executing the BL path with ID $38$, $66\%$ of the times the program executes path $86$, and $28\%$ of the times BL path $87$. When $86$ follows $38$, $100\%$ of the times the control flow takes the path $\langle 86, 86, 86, 755\rangle$, which spans four loop iterations and may be successfully unrolled to perform trace scheduling. Interestingly, sequence $\langle 38, 86, 86, 86, 755, 38, 86, 86, 86, 755\rangle$ of 11 BL path IDs, highlighted in Figure~\ref{fig:eclipse-example}, accounts for more than $50\%$ of all execution of the first BL path in the sequence, showing that sequence $\langle 38, 86, 86, 86, 755\rangle$ is likely to be repeated consecutively more than once. 


\subsection{Algorithms}

In this section, we show how to efficiently construct a \kipf\ profile starting from a stream of BL path IDs. 
The main idea is to construct an intermediate data structure that can be updated quickly, and then convert this data structure into a \kipf\ more efficiently when the stream is over. As intermediate data structure, we use a variant of the {\em $k$-slab forest} (\ksf) introduced in~\cite{Ausiello:2012:KCP:2384616.2384679}, which we adapt to our context as follows:

\begin{definition}[$k$-slab forest]
\label{def:k-sf}
Let $k\ge 2$ and let $s_1, s_2, s_3, \ldots,$ $s_m$ be the subsequences of $\Sigma$ obtained by: (1) splitting $\Sigma$ at $*$ markers, (2) removing the markers, and (3) cutting the remaining subsequences every $k-1$ consecutive items. The {\em $k$-slab forest} (\ksf) of $\Sigma$ is defined as $\ksf={\cal F}($list of all prefixes of $s_1\cdot s_2$ and all prefixes of length $\ge k$ of $s_i\cdot s_{i+1}$, $\forall i\in [2,m-1])$, where $s_i\cdot s_{i+1}$ denotes the concatenation of $s_i$ and $s_{i+1}$.
\end{definition}

\noindent By Definition~\ref{def:k-sf}, since each $s_i$ has length up to $k-1$, then a \ksf\ has at most $2k-2$ levels and depth $2k-3$.

\begin{figure}
\centerline{\includegraphics[width=0.99\columnwidth]{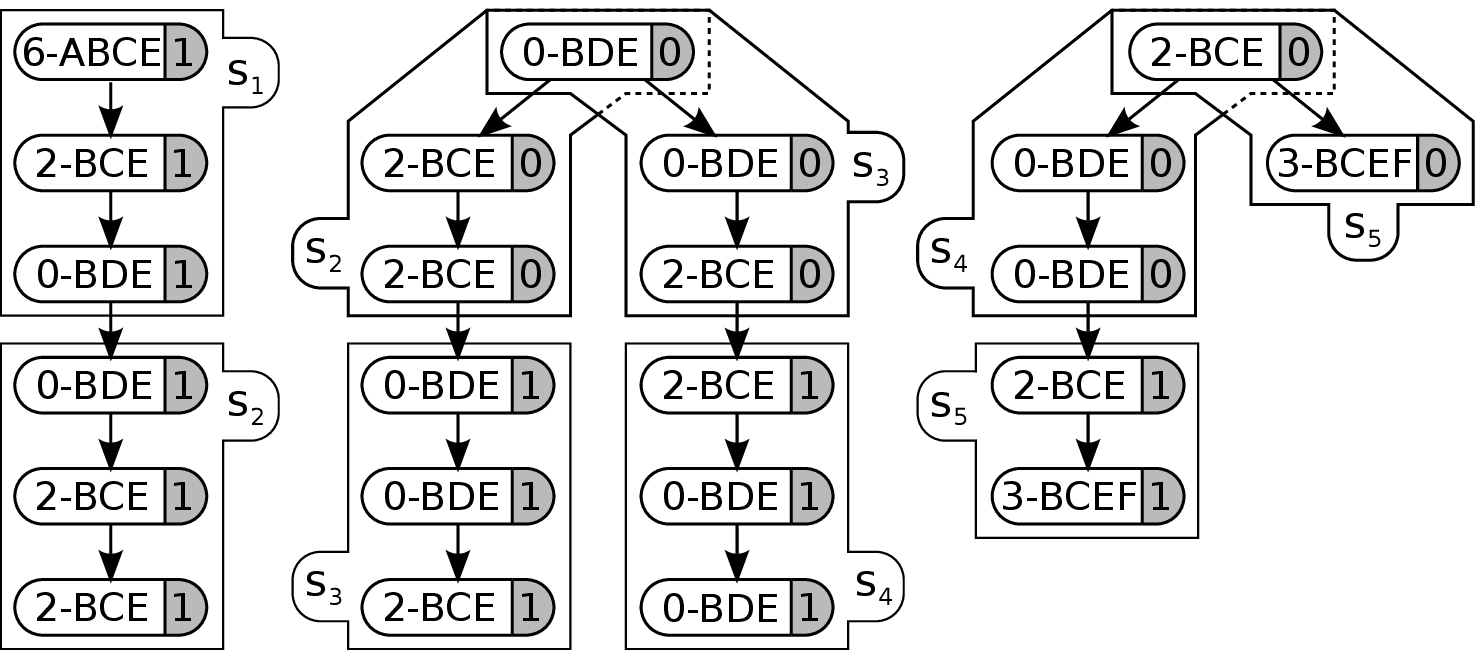}}
\caption{4-SF resulting from the execution trace of Figure~\ref{fig:example}.}
\label{fig:ksf}
\end{figure}

\paragraph{Example 3.} Let us consider again the example given in Figure~\ref{fig:example}. For $k=4$, we break the stream into maximal subsequences of up to $k-1=3$ consecutive BL path IDs: $$\Sigma=\langle *, \underbrace{\framebox{6, 2, 0}}_{s_1}, \underbrace{\framebox{0, 2, 2}}_{s_2}, \underbrace{\framebox{0, 0, 2}}_{s_3}, \underbrace{\framebox{2, 0, 0}}_{s_4}, \underbrace{\framebox{2, 3}}_{s_5}\rangle.$$

\noindent The $4$-SF of $\Sigma$, defined in terms of $s1, \ldots, s_5$, is shown in Figure~\ref{fig:ksf}. The forest is obtained as ${\cal F}(L)$, where $L=\langle$ $
\langle 6\rangle, 
\langle 6, 2\rangle, 
\langle 6, 2, 0\rangle, 
\langle 6, 2, 0, 0\rangle, 
\langle 6, 2, 0, 0, 2\rangle, 
\langle 6, 2, 0, 0, 2, 2\rangle,$
$
\langle 0, 2, 2, 0\rangle, 
\langle 0, 2, 2, 0, 0\rangle, 
\langle 0, 2, 2, 0, 0, 2\rangle,$
$
\langle 0, 0, 2, 2\rangle,$ 
$\langle 0, 0,2,$ $2, 0\rangle,$
$\langle 0, 0, 2, 2, 0, 0\rangle,$
$
\langle 2, 0, 0, 2\rangle,
\langle 2, 0, 0, 2, 3\rangle
\rangle$.



\begin{figure}
\begin{small}
\noindent {\bf procedure} {\tt process\_bl\_path\_id}$(r)$:
\begin{algorithmic}[1]
\IF {$r=*$}
\STATE $n\gets 0$
\STATE $\tau\gets{\tt null}$
\RETURN
\ENDIF
\IF  {$n$ {\bf mod} $(k-1)=0$}
\STATE $\beta\leftarrow \tau$ 
\STATE $\tau\,\leftarrow\,${\tt find$(R,r)$}
\IF {$\tau = \texttt{null}$ }
\STATE add root $\tau$ with $\ell(\tau)=r$ and $c(\tau)=0$ to \ksf\ and $R$
\ENDIF
\ELSE
\STATE find child $\omega$ of node $\tau$ with label $\ell(\omega)=r$
\IF {$\omega$ = \texttt{null} }
\STATE add node $\omega$ with $\ell(\omega)=r$ and $c(\omega)=0$ to \ksf\
\STATE add arc $(\tau,\omega)$ to \ksf\
\ENDIF
\STATE $\tau\leftarrow \omega$
\ENDIF
\IF {$\beta\neq{\tt null}$}
\STATE find child $\upsilon$ of node $\beta$ with label $\ell(\upsilon)=r$ 
\IF {$\upsilon = \texttt{null}$}
\STATE add node $\upsilon$ with $\ell(\upsilon)=r$ and $c(\upsilon)=0$ to \ksf\
\STATE add arc $(\beta,\upsilon)$ to \ksf\
\ENDIF
\STATE $\beta\leftarrow \upsilon$
\STATE $c(\beta)\gets c(\beta)+1$
\ELSE
\STATE $c(\tau)\gets c(\tau)+1$
\ENDIF
\STATE $n\gets n+1$
\end{algorithmic}
\end{small}
\smallskip
\caption{Streaming algorithm for \ksf\ construction.}
\label{fig:k-sf-algorithm}
\end{figure}

\paragraph{k-SF Construction Algorithm.} Given a stream $\Sigma$ formed by $*$ markers and BL path IDs, the \ksf\ of $\Sigma$ can be constructed by calling the procedure {\tt process\_bl\_path\_id}$(r)$ shown in Figure~\ref{fig:k-sf-algorithm} on each item $r$ of $\Sigma$. The streaming algorithm, which is a variant of the \ksf\ construction algoritm given in~\cite{Ausiello:2012:KCP:2384616.2384679} for the different setting of bounded-length calling contexts, keeps the following information:

\begin{itemize}
\item a hash table $R$, initially empty, containing pointers to the roots of trees in the \ksf, hashed by node labels; since no two roots have the same label, the lookup operation {\tt find}$(R, r)$ returns the pointer to the root containing label $r$, or {\tt null} if no such root exists;
\item a variable $n$ that counts the number of BL path IDs processed since the last $*$ marker;
\item a variable $\tau$ (top) that points either to {\tt null}, or to the current \ksf\ node in the upper part of the forest (levels 0 through $k-2$);
\item a variable $\beta$ (bottom) that points either to {\tt null}, or to the current \ksf\ node in the lower part of the forest (levels $k-1$ through $2k-3$).
\end{itemize}

\noindent The main idea of the algorithm is to progressively add new paths to an initially empty \ksf. The path formed by the first $k-1$ items since the last $*$ marker is added to one tree of the upper part of the forest. Each later item $r$ is added at up to two different locations of the \ksf: one in the upper part of the forest (lines 13--17) as a child of node $\tau$ (if no child of $\tau$ labeled with $r$ already exists), and the other one in the lower part of the forest (lines 21--25) as a child of node $\beta$ (if no child of $\beta$ labeled with $r$ already exists). Counters of processed nodes already containing $r$ are incremented by one (either line 27 or line 29). Both $\tau$ and $\beta$ are updated to point to the child labeled with $r$ (lines 18 and 26, respectively). The running time of the algorithm is dominated by lines 8 and 10 (hash table accesses), and by lines 13 and 21 (node children scan). Assuming that operations on $R$ require constant time, the per-item processing time is $O(\delta)$, where $\delta$ is the maximum degree of a node in the \ksf. Our experiments revealed that $\delta$ is on average a typically small constant value.

\begin{figure}
\begin{small}
\noindent {\bf procedure} {\tt make\_k\_ipf}$()$:
\begin{algorithmic}[1]
\STATE $I\gets\emptyset$
\FOR {each node $\rho\in\ksf$}
\IF {$\ell(\rho)\not\in I$}
\STATE add $\ell(\rho)$ to $I$ and let $s(\ell(\rho))\gets\emptyset$
\ENDIF
\STATE add $\rho$ to $s(\ell(\rho))$
\ENDFOR
\STATE let the \kipf\ be formed by a dummy root $\phi$
\FOR {each  $r\in I$}
\FOR {each  $\rho\in s(r)$}
\STATE {\tt join\_subtree}$(\rho,\phi,k)$
\ENDFOR
\ENDFOR
\STATE remove dummy root $\phi$ from the \kipf\
\end{algorithmic}
\medskip
\noindent {\bf procedure} {\tt join\_subtree}$(\rho,\gamma,d)$:
\begin{algorithmic}[1]
\STATE $\delta\gets$ child of $\gamma$ in the \kipf\ s.t. $\ell(\delta)=\ell(\rho)$
\IF {$\delta={\tt null}$}
\STATE add new node $\delta$ as a child of $\gamma$ in the \kipf\
\STATE $\ell(\delta)\gets\ell(\rho)$ and $c(\delta)\gets c(\rho)$
\ELSE
\STATE $c(\delta)\gets c(\delta)+c(\rho)$
\ENDIF
\IF {$d>0$}
\FOR {each child $\sigma$ of $\rho$ in the \ksf}
\STATE {\tt join\_subtree}$(\sigma, \delta, d-1)$
\ENDFOR
\ENDIF
\end{algorithmic}
\end{small}
\smallskip
\caption{Algorithm for converting a \ksf\ into a \kipf.}
\label{fig:ksf2kipf-algo}
\end{figure}

\paragraph{\ksf\ to \kipf\ Conversion.} Once the stream $\Sigma$ is over, i.e., the profiled thread has terminated, we convert the \ksf\ into a \kipf\ using the procedure {\tt make\_k\_ipf} shown in Figure~\ref{fig:ksf2kipf-algo}. 
The algorithm creates a set $I$ of all distinct path IDs that occur in the \ksf\ and for each $r$ in $I$ builds a set $s(r)$ containing all nodes $\rho$ of the \ksf\ labeled with $r$ (lines 2--7). To build the \kipf, the algorithm lists each distinct path ID $r$ and joins to the \kipf\ all subtrees of depth up to $k-1$ rooted at a node in $s(r)$ in the \ksf, as children of a dummy root, which is added for the sake of convenience and then removed. The join operation is specified by procedure {\tt join\_subtree}, which performs a traversal of a subtree of the \ksf\ of depth less than $k$ and adds nodes to \kipf\ so that all labeled paths in the subtree appear in the \kipf\ as well, but only once. Path counters in the \ksf\ are accumulated in the corresponding nodes of the \kipf\ to keep track of the number of times each distinct path consisting of the concatenation of up to $k$ BL paths was taken by the profiled program.\rem{Costo conversione.}




\section{Implementation}
\label{se:implementation}

In this section we describe the implementation of our profiler, which we call k-BLPP, in the Jikes Research Virtual Machine~\cite{Jikes}.

\subsection{Adaptive Compilation}

The Jikes RVM is a high performance metacircular virtual machine: unlike most others JVMs, it is written in Java. Jikes RVM does not include an interpreter: all bytecode must be first translated into native machine code. The unit of compilation is the method, and methods are compiled lazily by a fast non-optimizing compiler -- the so-called {\em baseline} compiler -- when they are first invoked by the program. As execution continues, the Adaptive Optimization System monitors program execution to detect program hot spots and selectively recompiles them with three increasing levels of optimization. Note that all modern production JVMs rely on some variant of selective optimizing compilation to target the subset of the hottest program methods where they are expected to yield the most benefits.

Recompilation is performed by the {\em optimizing} compiler, that generates higher-quality code but at a significantly larger cost than the baseline compiler. Since Jikes RVM quickly recompiles frequently executed methods, we implemented k-BLPP in the optimizing compiler only.

\subsection{Inserting Instrumentation on Edges}
\kblpp\ adds instrumentation to hot methods in three passes:
\begin{enumerate}
 \item building the DAG representation;
 \item assigning values to edges;
 \item adding instrumentation to edges.
\end{enumerate}

\noindent \kblpp\ adopts the {\em smart path numbering} algorithm proposed by Bond and McKinley~\cite{BM05b} to improve performance by placing instrumentation on cold edges. In particular, line 6 of the canonical Ball-Larus path numbering algorithm shown in Figure~\ref{fig:bl-numbering} is modified such that outgoing edges are picked in decreasing order of execution frequency. For each basic block edges are sorted using existing edge profiling information collected by the baseline compiler, thus allowing us to assign zero to the hottest edge so that \kblpp\ does not place any instrumentation on it.

During compilation, the Jikes RVM generates {\em yield points}, which are program points where the running thread determines if it should yield to another thread. Since JVMs need to gain control of threads quickly, compilers insert yield points in method prologues, loop headers, and method epilogues. We modified the optimizing compiler to also store the path profiling probe on loop headers and method epilogues. Ending paths at loop headers rather than back edges causes a path that traverse a header to be split into two paths: this difference from canonical Ball-Larus path profiling is minor because it only affects the first path through a loop~\cite{BM05}.

Note that optimizing compilers do not always insert yield points: this occurs when a method either does not contain branches (hence its profile is trivial) or is marked as uninterruptible. The second case occurs in internal Jikes RVM methods only; the compiler occasionally inlines such a method into an application method, and this might result in a loss of information only when the execution reaches a loop header contained in the inlined method. However, this loss of information appears to be negligible~\cite{BM05}.

\subsection{Path Profiling}
The \ksf\ construction algorithm described in Section~\ref{sse:kIterationsPP} is implemented using a standard first-child, next-sibling representation for nodes: this representation is very space-efficient, while experimental results show that the average degree of a node is usually low.

Tree roots are stored and accessed through an efficient stripped-down implementation of a hash map, using the pair represented by the Ball-Larus path ID and the unique identifier associated to the current routine as key. Note that this map is typically smaller than a map required by a traditional BLPP profiler, since tree roots represent only a fraction of the distinct path IDs encountered during the execution. Consider, for instance, a routine with $N$ acyclic paths whose control flow graph contains a common and unique binary branch before the first cycle is entered: since cyclic paths are truncated on loop headers, only two distinct path IDs can appear as a tree root in the hash map, while the remaining $N-2$ paths can appear only inside non-root nodes.


\begin{figure*}[t]
\centerline{\includegraphics[width=0.99\textwidth]{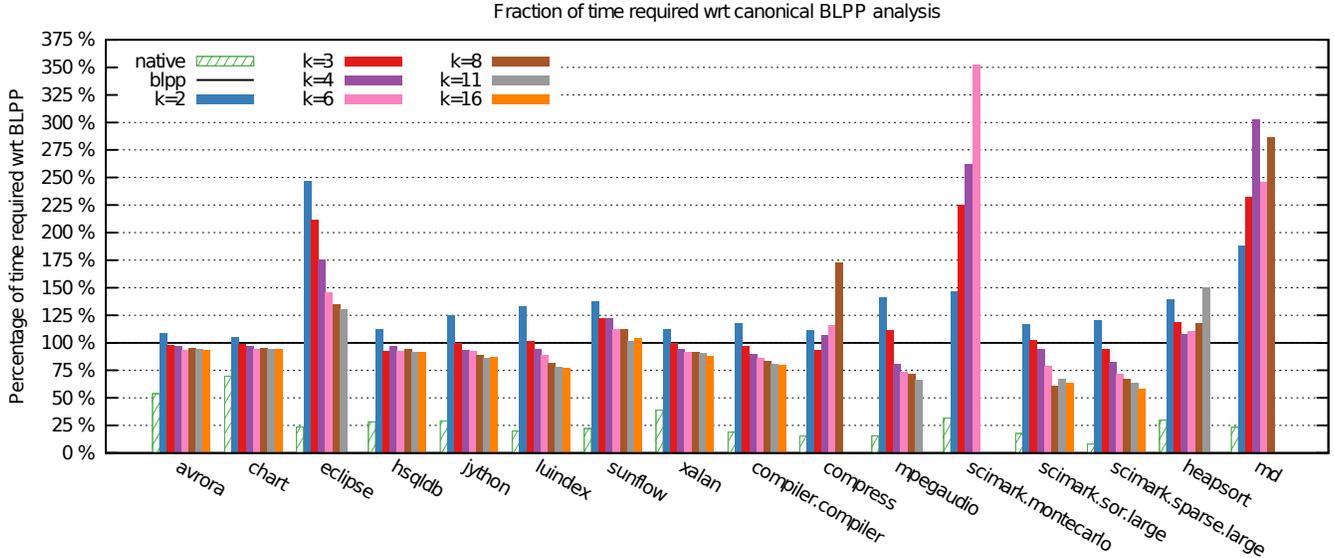}}
\vspace{-2mm}
\caption{Performance of \kblpp\ relative to \blpp.}
\label{fig:benchmarks-time}
\end{figure*}

\begin{figure*}[t]
\centerline{\includegraphics[width=0.99\textwidth]{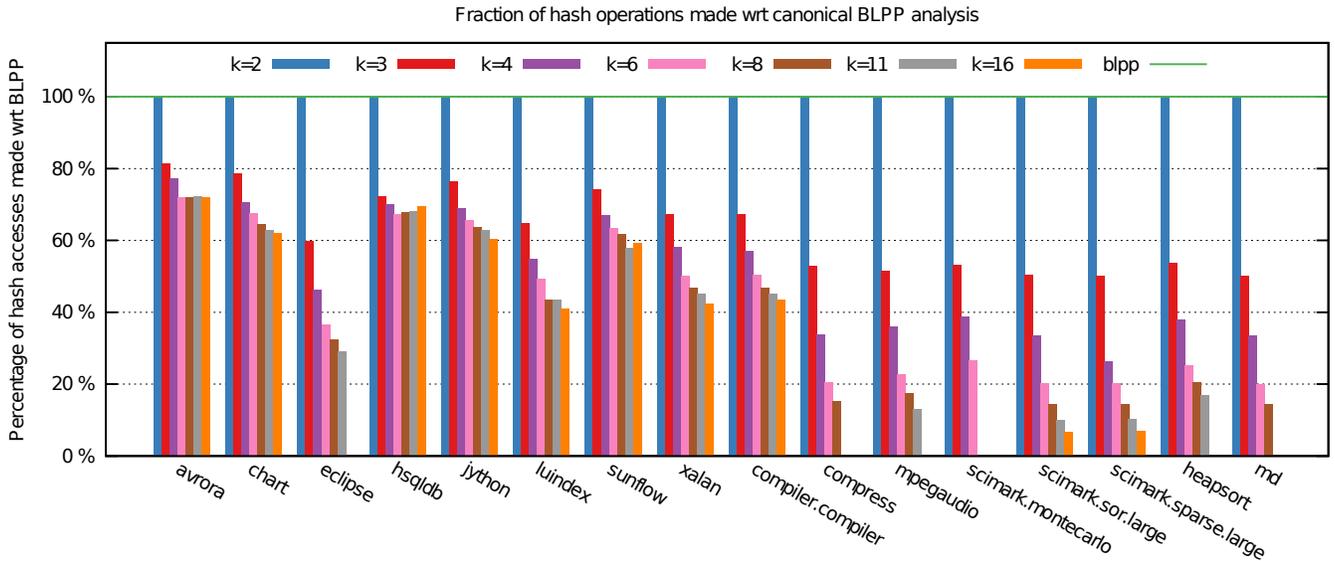}}
\vspace{-2mm}
\caption{Number of hash table operations performed by \kblpp\ relative to \blpp.}
\label{fig:benchmarks-hash}
\end{figure*}

\section{Experimental Evaluation}
\label{se:experiments}

In this section we report the result of an extensive experimental evaluation of our approach. The goal is to assess the performance of our profiler compared to previous approaches and to study properties of path profiles that span multiple iterations for several representative benchmarks.


\subsection{Experimental Setup}

\paragraph{Bechmarks.} We evaluated k-BLPP against a variety of prominent benchmarks drawn from three suites. The {\tt DaCapo} suite~\cite{DaCapo} consists of a set of open source, real-world applications with non-trivial memory loads. We use the superset of all benchmarks from {\tt DaCapo} releases 2006-MR2 and 9.12 that can run successfully with Jikes RVM, using the largest available workload for each benchmark. The {\tt SPEC} suite focuses on the performance of the hardware processor and memory subsystem when executing common general purpose application computations\footnote{Unfortunately, only a few benchmarks from SPEC JVM2008 can run successfully with Jikes RVM due to limitations of the GNU classpath.}. Finally, we chose two memory-intensive benchmarks from the {\tt Java Grande} $2.0$ suite~\cite{JavaGrande} to further evaluate the performance of k-BLPP.


\paragraph{Compared Codes.} In our experiments, we analyzed the native (uninstrumented) version of each benchmark and its instrumented counterparts, comparing \kblpp\ for different values of $k$ (2, 3, 4, 6, 8, 11, 16) with an updated version of the \blpp\ profiler developed by Bond~\cite{BM05,JikesResearchArchive}, which implements the Ball-Larus acyclic path profiling technique.


\paragraph{Platform.} Our experiments were performed on a 2.53GHz Intel Core2 Duo T9400 with 128KB of L1 data cache, 6MB of L2 cache, and 4 GB of main memory DDR3 1066, running Ubuntu 12.10, Linux Kernel 3.5.0, 32 bit. We ran all of the benchmarks on Jikes RVM 3.1.3 (default {\tt production} build) using a single core and a maximum heap size equal to half of the amount of physical memory.

\paragraph{Metrics.} We considered a variety of metrics, including wall-clock time, number of operations per second performed by the profiled program, number of hash table operations, data structure size (e.g., number of hash table items for \blpp\ and number of \ksf\ nodes for \kblpp), and statistics such as average node degree of the \ksf\ and the \kipf\ and average depth of \kipf\ leaves. To interpret our results, we also ``profiled our profiler'' by collecting hardware performance counters with {\tt perf}~\cite{perf}, including L1 and L2 cache miss rate, branch mispredictions, and cycles per instruction (CPI).

\paragraph{Methodology.} For each benchmark/profiler combination, we performed at least 7 trials, each preceded by a warmup execution, and computed the arithmetic mean. We monitored variance, increasing the number of trials for problematic benchmarks. Performance measurements were collected on a machine with negligible background activity. 


\begin{figure*}
\noindent
\begin{tabular}{ccc}
\includegraphics[width=0.315\textwidth]{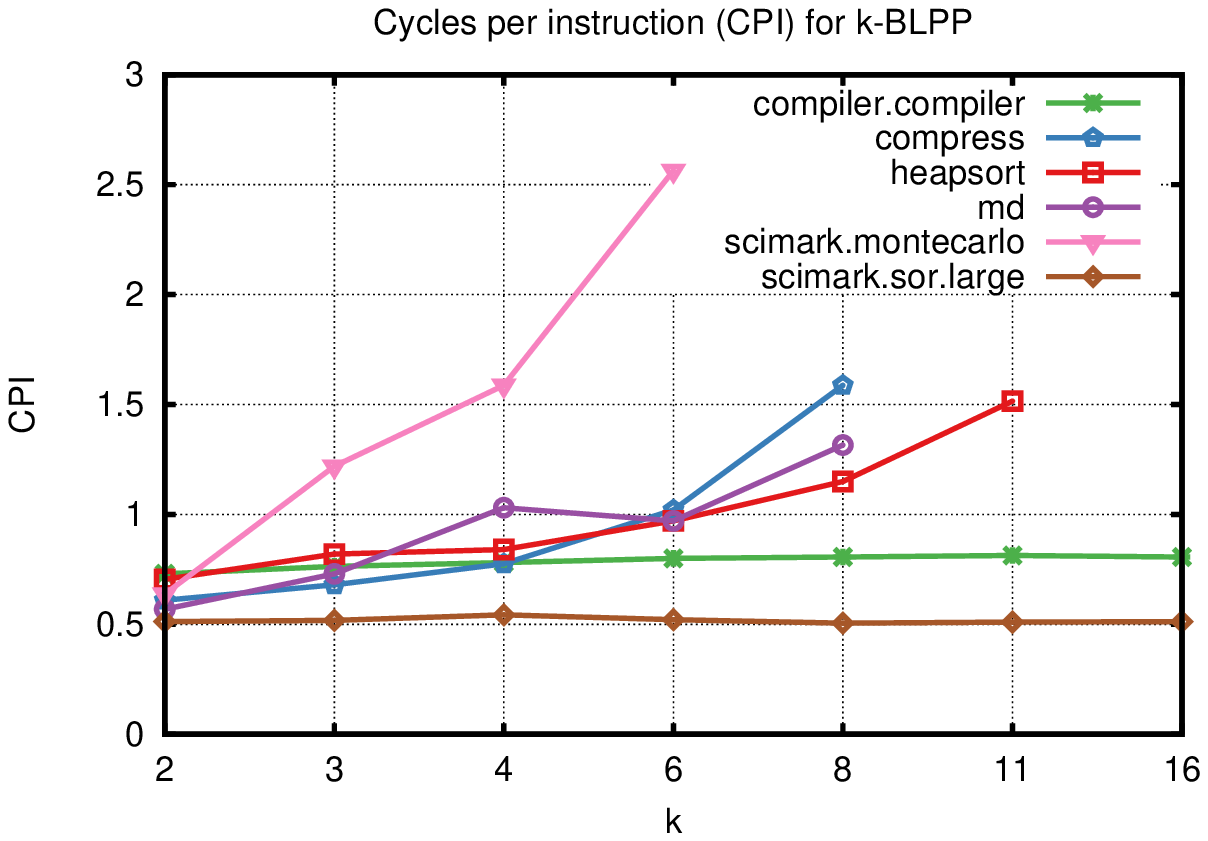} & \includegraphics[width=0.315\textwidth]{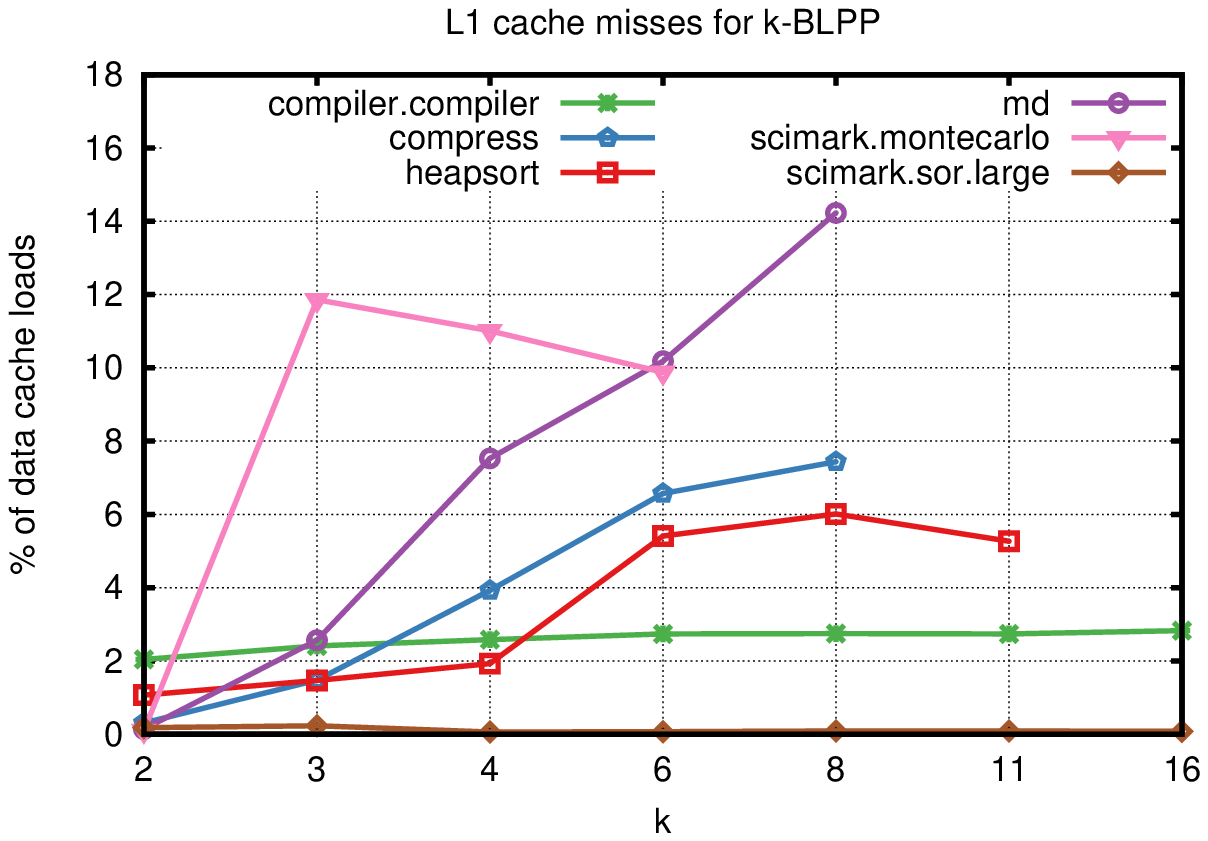} & \includegraphics[width=0.315\textwidth]{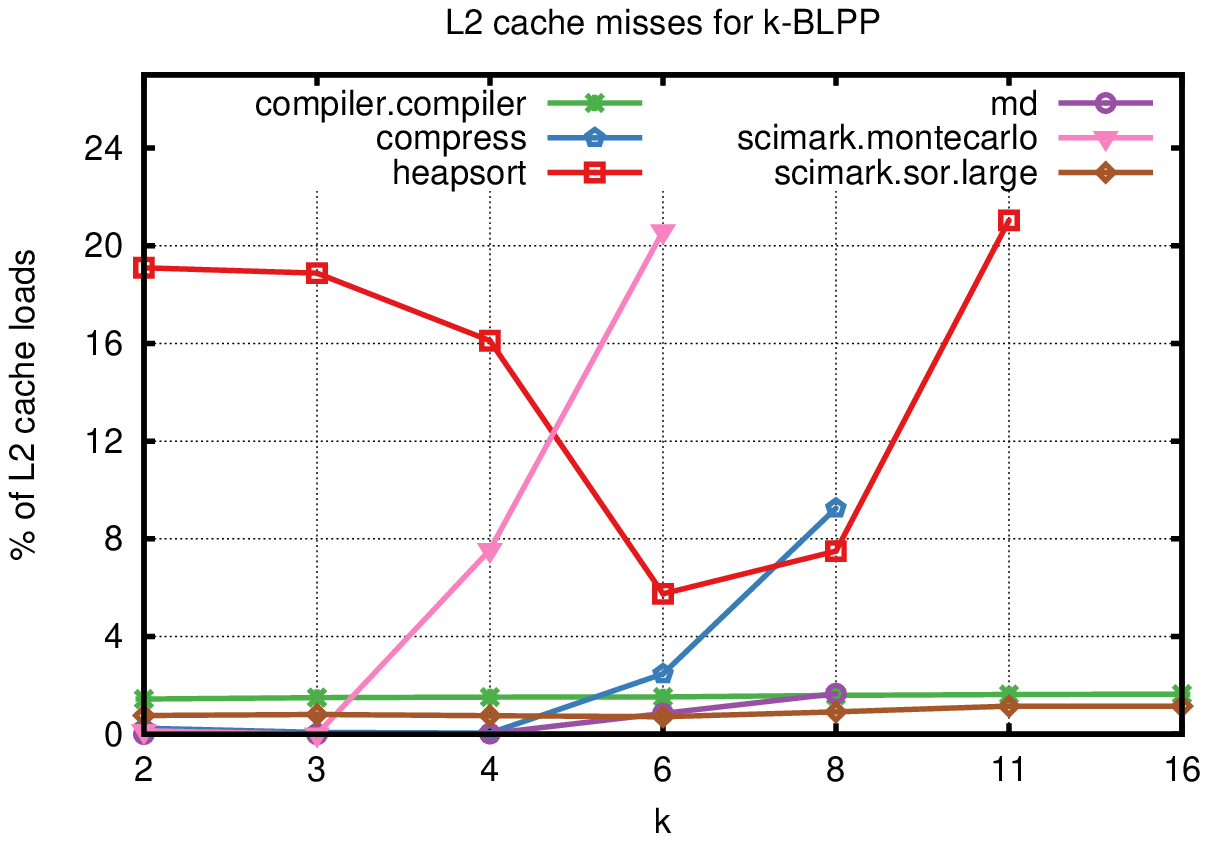} \\
(a) & (b) & (c)
\end{tabular}
\vspace{-1mm}
\caption{Hardware performance counters for \kblpp: (a) cycles per instruction, (b) L1 cache miss rate, (c) L2 cache miss rate.}
\label{fig:cpi}
\end{figure*}

\begin{figure*}
\centerline{\includegraphics[width=0.99\textwidth]{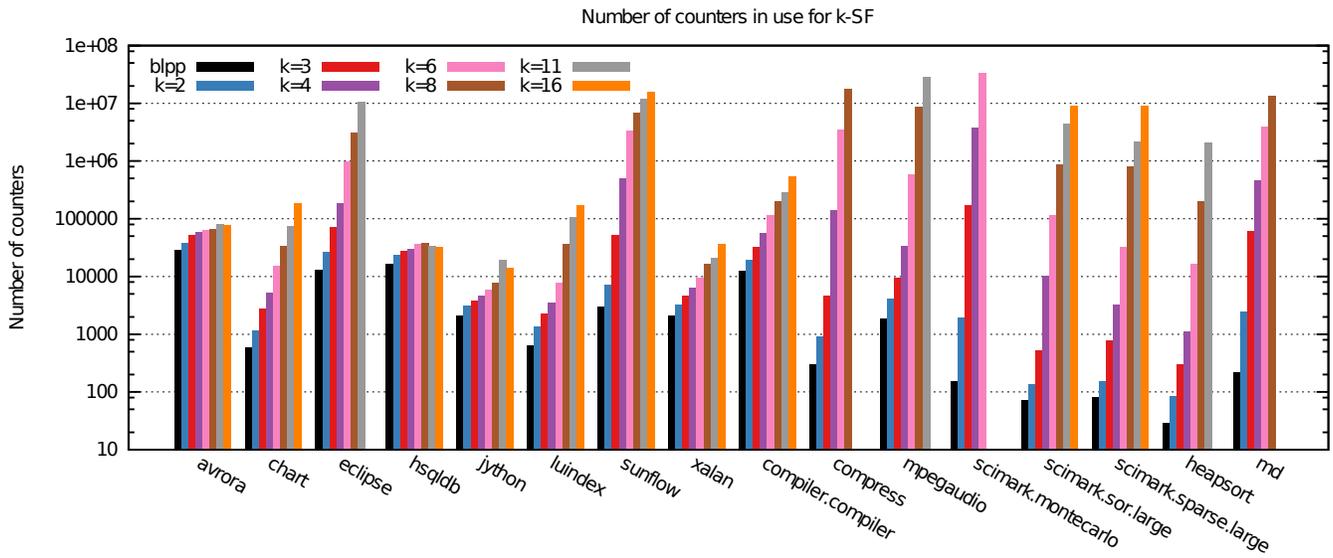}}
\vspace{-2mm}
\caption{Space requirements: number of hash table entries in \blpp\ and number of nodes in the \ksf.}
\label{fig:benchmarks-space}
\end{figure*}

\begin{figure*}
\centerline{\includegraphics[width=0.99\textwidth]{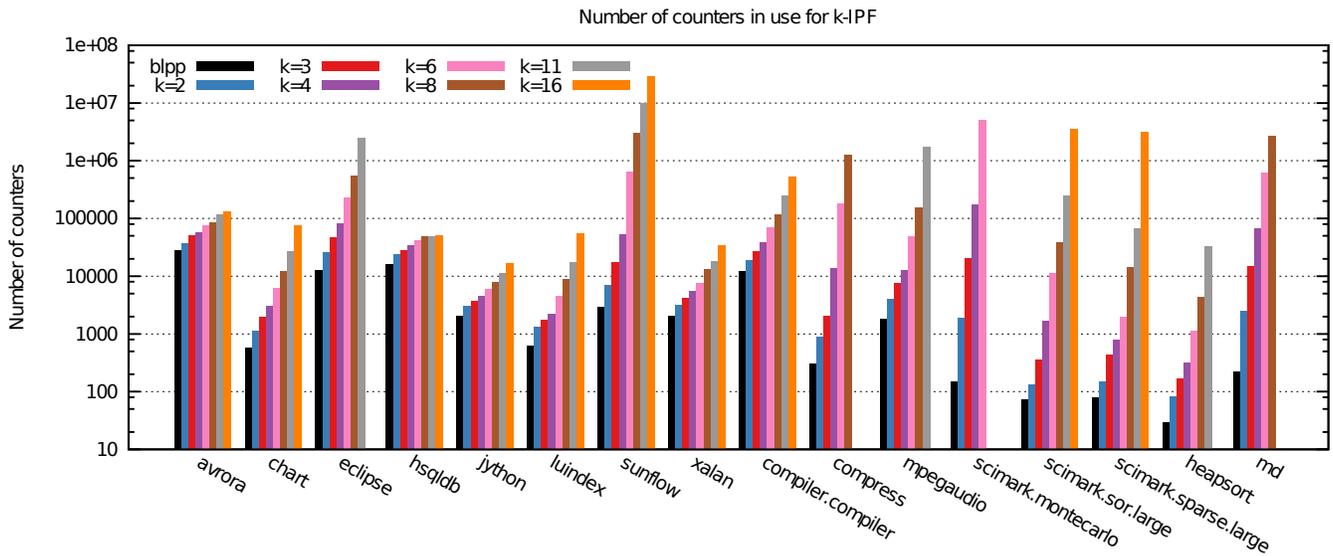}}
\vspace{-2mm}
\caption{Number of paths profiled by \blpp\ and \kblpp.}
\label{fig:benchmarks-ipf-nodes}
\end{figure*}

\begin{figure*}
\centerline{\includegraphics[width=0.99\textwidth]{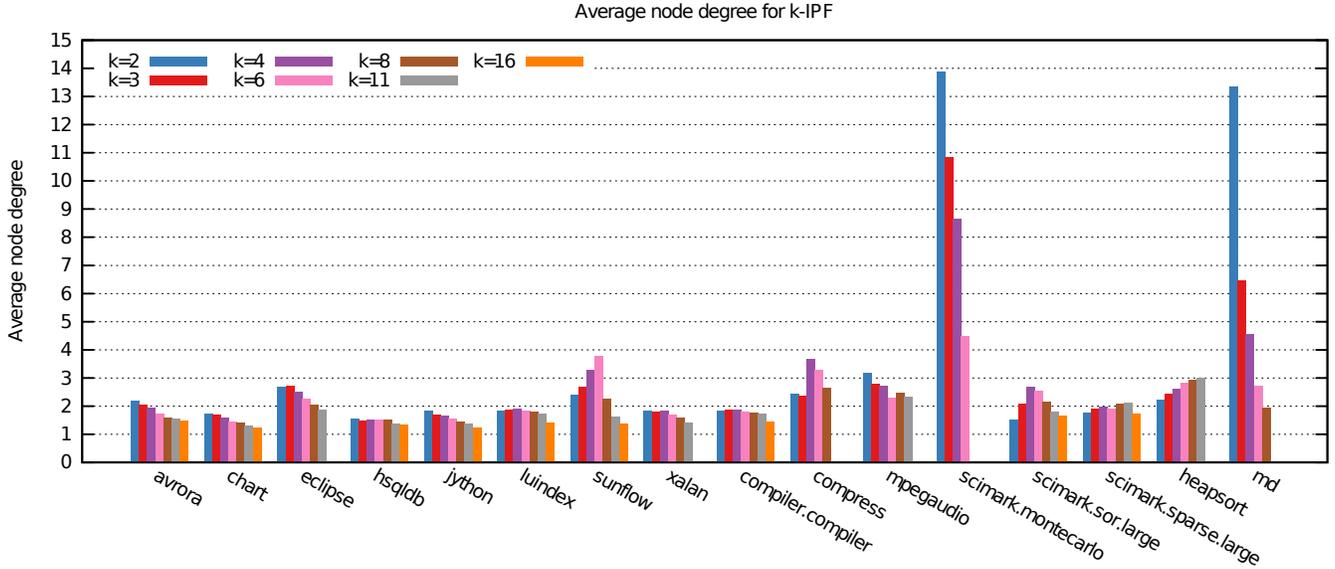}}
\vspace{-2mm}
\caption{Average degree of \kipf\ internal nodes.}
\label{fig:benchmarks-ipf-node-degr}
\end{figure*}

\begin{figure*}
\centerline{\includegraphics[width=0.99\textwidth]{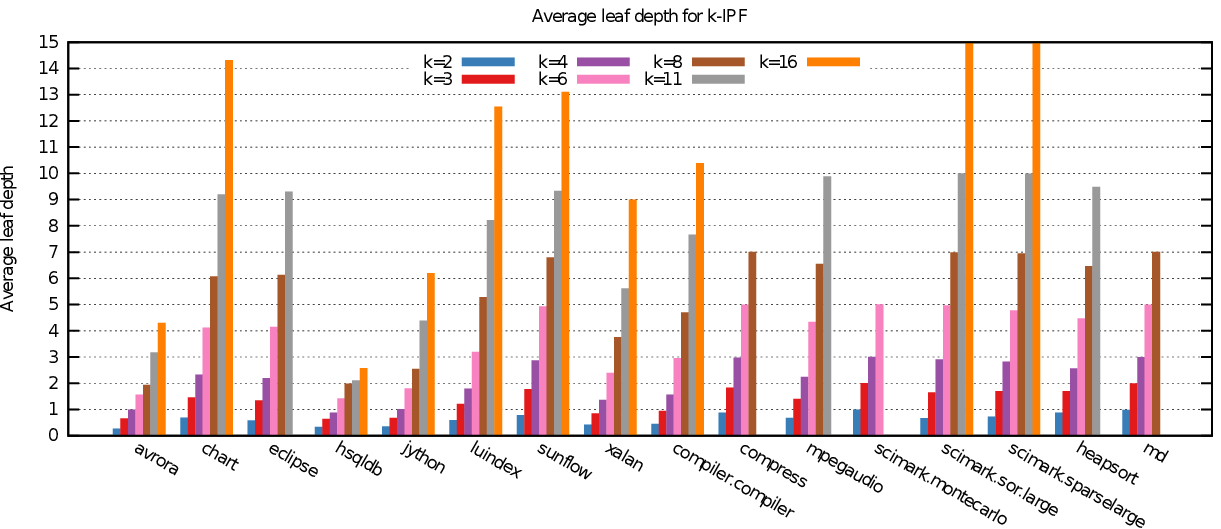}}
\vspace{-2mm}
\caption{Average depth of \kipf\ leaves.}
\label{fig:benchmarks-ipf-leaves-depth}
\end{figure*}

\subsection{Experimental Results}

\paragraph{Performance overhead.} In Figure~\ref{fig:benchmarks-time} we report for each benchmark the profiling overhead of \kblpp\ relative to \blpp. The chart shows that for 12 out of 16 benchmarks the overhead decreases for increasing values of $k$, providing up to almost 50\% improvements over \blpp. This is explained by the fact that hash table accesses are performed by {\tt process\_bl\_path\_id} every $k-1$ items read from the input stream between two consecutive routine entry events (lines 8 and 10 in Figure~\ref{fig:k-sf-algorithm}). As a consequence, the number of hash table operations for each routine call is $O(1+N/(k-1))$, where $N$ is the total length of the path taken during the invocation. In Figure~\ref{fig:benchmarks-hash} we report the measured number of hash table accesses for our experiments, which decreases as predicted on all benchmarks with intense loop iteration activity. Notice that, not only \kblpp\ performs fewer hash table operations, but since only a subset of BL path IDs are inserted, the table is also smaller yielding further performance improvements. For codes such as {\tt avrora} and {\tt hsqldb}, which perform on average a small number of iterations, increasing $k$ beyond this number does not yield any benefit.


On {\tt eclipse}, \kblpp\ gets faster as $k$ increases, but differently from all other benchmarks in this class, remains slower than \blpp. The reason is that, due to structural properties of the benchmark, the average number of node scans at lines 13 and 21 of {\tt process\_bl\_path\_id} is rather high (58.8 for $k=2$ down to 10.3 for $k=16$). In contrast, the average degree of internal nodes of the \ksf\ is small (2.6 for $k=2$ decreasing to 1.3 for $k=16$), hence there is intense activity on nodes with a high number of siblings. No other benchmark exhibited this extreme behavior. We expect that a more efficient implementation of {\tt process\_bl\_path\_id}, e.g., by adaptively moving hot children to the front of the list, could reduce the scanning overhead for this kind of worst-case benchmarks as well.

Benchmarks {\tt compress}, {\tt scimark.monte\_carlo}, {\tt heap- sort}, and {\tt md} made an exception to the general trend we observed, with performance overhead increasing, rather than decreasing, with $k$. To justify this behavior, we collected and analyzed several hardware performance counters and noticed that on these benchmarks our \kblpp\ implementation suffers from increased CPI for higher values of $k$. Figure~\ref{fig:cpi} (a) shows this phenomenon, comparing the four outliers with other benchmarks in our suite. By analyzing L1 and L2 cache miss rates, reported in Figure~\ref{fig:cpi} (b) and Figure~\ref{fig:cpi} (c), we noticed that performance degrades due to poor memory access locality. We believe this to be an issue of our current implementation of \kblpp, in which we did not make any effort aimed at improving cache efficiency, rather than a limitation of the general approach we propose.

\paragraph{Space Usage.} Figure~\ref{fig:benchmarks-space} compares the space requirements of \blpp\ and \kblpp\ for different values of $k$. The chart reports the total number of items stored in the hash table by \blpp\ and the number of nodes in the \ksf. Since both \blpp\ and \kblpp\ exhaustively encode exact counters for all distinct taken paths of bounded length, space depends on intrinsic structural properties of the benchmark. Programs with intense loop iteration activity are characterized by substantially higher space requirements by \kblpp, which collects profiles containing up to several millions of paths. Notice that on some benchmarks we ran out of memory for large values of $k$, hence some bars in the charts we report in this section are missing. In Figure~\ref{fig:benchmarks-ipf-nodes} we report the number of nodes in the \kipf, which corresponds to the number of paths profiled by \kblpp. Notice that, since a path may be represented more than once in the \ksf, the \kipf\ represents a more compact version of the \ksf.

\begin{table*}
\begin{center}
\begin{footnotesize}
\begin{tabular}{ l|cccccc }
  \hline                        
  Techniques & \blpp~\cite{BL96} & SPP~\cite{AH02}, TPP~\cite{JBZ04}, PPP~\cite{BM05b} & Tallam {\em et al.}~\cite{Tallam:2004:EPP:977395.977659} &  Roy {\em et al.}~\cite{SS09} & Li {\em et al.}~\cite{Li20121558} & This paper\\ \hline\hline  
  {\multirow{2}{*}{Goals} } & profiling & reduce \blpp\ & profiling & profiling & profiling & profiling \\ 
   & acyclic paths & overhead & 2-iteration paths & cyclic paths & cyclic paths & cyclic paths \\ 
  Profiled paths & BL paths & subset of BL paths & overlapping paths & $k$-iteration paths & finite-length paths & $k$-iteration paths \\ 
 Avg. cost w.r.t. \blpp\ & - & smaller & larger & larger & larger & smaller \\ 
  Variables & 1 & 1 or more & many & many & 1 or more & 1 \\ 
  Full accuracy & $\surd$ & $\surd$ & & $\surd$ & $\surd$ & $\surd$ \\ 
  Inter-procedural &  & & $\surd$ & & & \\ \hline
\end{tabular}
\end{footnotesize}
\end{center}
\vspace{-3mm}
\caption{Comparison of different path profiling techniques.}
\label{tab:related}
\end{table*}

\paragraph{Structural Properties of Collected Profiles.} As a final experiment, we measured structural properties of the \kipf\ such as average degree of internal nodes (Figure~\ref{fig:benchmarks-ipf-node-degr}) and the average leaf depth (Figure~\ref{fig:benchmarks-ipf-leaves-depth}). Our tests reveal that the average node degree generally decreases with $k$, showing that similar patterns tend to appear frequently across different iterations. Some benchmarks, however, such as {\tt sunflow} and {\tt heapsort} exhibit a larger variety of path ramifications, witnessed by increasing node degrees at deeper levels of the \kipf. The average leaf depth allows it to characterize the loop iteration activity of different benchmarks. Notice that some benchmarks, such as {\tt avrora} and {\tt hsqldb}, have short cycles. Hence, by increasing $k$ beyond the maximum cycle length, \kblpp\ does not collect any additional information.

\paragraph{Discussion.} 
From our experiments, we could draw two main conclusions:


\begin{enumerate}
\item Using tree-based data structures to represent intraprocedural control flow allows it to substantially reduce the performance overhead of path profiling by decreasing the number of hash operations, which also operate on smaller tables. This approach yields the first profiler that can handle loops that extend across multiple loop iterations faster than the general Ball-Larus technique based on hash tables for maintaining path frequency counters, while collecting at the same time significantly more informative profiles. We observed that, due to limitations of our current implementation of \kblpp\ such as lack of cache friendliness for some worst-case scenarios, on a few outliers our profiler was slower than Ball-Larus, with a peak of 3.5x slowdown on one benchmark.
\item Since the number of profiled paths in the control flow graph typically grows exponentially for increasing values of $k$, space usage can become prohibitive if paths spanning many loop iterations have to be exhaustively profiled. We noticed, however, that most long paths have small frequency counters, and are therefore 
uninteresting for identifying optimization opportunities. Hence, a useful addition to our method, which we do not address in this work, would be to prune cold nodes on-the-fly from the \ksf, keeping information for hot paths only.
\end{enumerate}


\section{Related Work}
\label{se:related}

\rem{Acyclic PP da articolo HCCT}

The seminal work of Ball and Larus~\cite{BL96} has spawned much research interest in the last 15 years, in particular on profiling acyclic paths with a lower overhead by using sampling techniques~\cite{BM05b,BM05} or choosing a subset of {\em interesting} paths~\cite{AH02,JBZ04,VNC07}. On the other hand, only a few works have dealt with cyclic paths profiling.

Tallam {\em et al.}~\cite{Tallam:2004:EPP:977395.977659} extend the Ball-Larus path numbering algorithm to record slightly longer paths across loop back edges and procedure boundaries. The extended Ball-Larus paths overlap and, in particular, are shorter than two iterations for paths that cross loop boundaries. These overlapping paths enable very precise estimation of frequencies of potentially much longer paths, with an average imprecision in estimated total flow of those paths ranging from $-4\%$ to $+8\%$. However, the average cost of collecting frequencies of overlapping paths is 4.2 times that of canonical \blpp\ on average.

Roy and Srikant~\cite{SS09} generalize the Ball-Larus algorithm for profiling $k$-iterations paths, showing that it is possible to number these paths efficiently using an inference phase to record executed backedges in order to differentiate cyclic paths. One problem with this approach is that, since the number of possible $k$-iteration paths grows exponentially with $k$, path IDs may overflow in practice already for small values of $k$ and very large hash tables may be required. In particular, their profiling procedure aborts if the number of static paths exceeds $60,000$, while this threshold is reached on several small benchmarks already for $k=3$~\cite{Li20121558}. This technique incurs a larger overhead than \blpp: in particular, the slowdown may grow to several times the \blpp-associated overhead as $k$ increases.


Li {\em et al.}~\cite{Li20121558} propose a new path encoding that does not rely on an inference phase to explicitly assign identifiers to all possible paths before the execution, yet ensuring that any finite-length acyclic or cyclic path has a unique ID. Their path numbering needs multiple variables to record probe values, which are computed by using addition and multiplication operations. Overflowing is handled by using {\em breakpoints} to store probe values: as a consequence, instead of a unique ID for each path, a unique series of breakpoints is assiged to each path. At the end of program's execution, the {\em backwalk} algorithm reconstructs the executed paths starting from breakpoints. This technique has been integrated with \blpp\ to reduce the execution overhead, resulting in a slowdown of about 2 times on average with respect to \blpp, but also showing significant performance loss (up to a 5.6 times growth) on tight loops. However, the experiments reported in~\cite{Li20121558} were performed on single methods of small Java programs, leaving further experiments on larger industry-strength benchmarks to future work.

The comparison of different path profiling techniques known in the literature with our approach is summarized in Table~\ref{tab:related}.

\section{Conclusions}
\label{se:conclusion}

In this paper we have presented a novel approach to cyclic path profiling, which combines the original Ball-Larus path numbering technique with a prefix tree data structure to keep track of concatenations of acyclic paths across multiple loop iterations. 
A large suite of experiments on a variety of prominent benchmarks shows that, not only our approach collects significantly more detailed profiles, but can also be  faster than the original Ball-Larus technique by reducing the number of hash table operations.


An interesting open question is how to use sampling-based approaches such as the one proposed by Bond and McKinley~\cite{BM05} to further reduce the path profiling overhead. We believe that the bursting technique, introduced by Zhuang {\em et al.}~\cite{ZSCC06} in the different scenario of calling context profiling could be successfully combined with our approach, allowing it to reduce the overhead while maintaining reasonable accuracy in mining hot paths.\rem{Guardare Arnold-Groove -- sampling points \& strides}

Another way to reduce the profiling overhead may be to exploit parallelism.\rem{citations to parallel analysis}
We note that our approach, which decouples path tracing from profiling using an intermediate data stream, is amenable to multi-core implementations by letting the profiled code and the analysis algorithm run on separate cores using shared buffers. A promising line of research is to explore how to partition the data structures so that portions of the stream buffer can be processed in parallel.

Finally, we observe that, since our approach is exhaustive and traces taken paths regardless of their hotness, it would be interesting to explore techniques for reducing space usage, by pruning cold branches of the \ksf\ on the fly to keep the memory footprint smaller, allowing it to deal with even longer paths.



\paragraph{Acknowledgments.} We are indebted to Michael Bond for several interesting discussions, for his invaluable support with his BLPP profiler, and for shedding light on some tricky aspects related to the Jikes RVM internals. We would also like to thank Erik Brangs for his help with the Jikes RVM and Jos\'e Sim\~ao for his hints in selecting an adequate set of benchmarks for the Jikes RVM.

\bibliographystyle{abbrvnat}
\softraggedright
\balance
\bibliography{DynamicAnalysis}

\end{document}